# Acoustic Anomaly Detection on UAM Propeller Defect

With Acoustic dataset for Crack of drone Propeller (ADCP)

**Authors List**

**1. Juho Lee**
Korea University, Seongbuk-gu, Seoul, South Korea
Captain, Korea Airforce, Student member, juho.lee927@gmail.com

**2. Donghyun Yoon**
Korea Advanced Institute of Science and Technology (KAIST), Yuseong-gu, Daejeon, South Korea
Ph.D. Student, KAIST, tloveu49@kaist.ac.kr

**3. Gumoon Jeong**
Korea Airforce Academy, Cheongju-si, Chungcheongbuk-do, South Korea
Lieutenant, Korea Airforce, entwurfjeong@gmail.com

**4.** (Corresponding author) **Hyeoncheol Kim**
Korea University, Seongbuk-gu, Seoul, South Korea
Professor, Korea University, hkim64@gmail.com

**Abstract**

**The imminent commercialization of UAM requires stable, AI-based maintenance systems to ensure safety for both passengers and pedestrians. This paper presents a methodology for non-destructively detecting cracks in UAM propellers using drone propeller sound datasets. Normal operating sounds were recorded, and abnormal sounds (categorized as ripped and broken) were differentiated by varying the microphone-propeller angle and throttle power. Our novel approach integrates FFT and STFT preprocessing techniques to capture both global frequency patterns and local time-frequency variations, thereby enhancing anomaly detection performance. The constructed Acoustic Dataset for Crack of Drone Propeller (ADCP) demonstrates the potential for detecting propeller cracks and lays the groundwork for future UAM maintenance applications.**

*keywords: UAM, Drone, Acoustic, Anomaly Detection, Maintenance*

# I. Introduction

In modern cities, where population density and traffic congestion continue to rise, the need for efficient urban transportation has become more critical than ever. UAM (Urban Air Mobility) is emerging as the next-generation solution to alleviate congestion by transporting passengers and cargo using aircraft that operate within city limits. Recent market analyses and governmental reports (Agency, 2021; Korea, 2021) highlight safety as a paramount concern, emphasizing that the rapid frequency and decentralized nature of UAM operations demand innovative maintenance strategies. Recent studies have also indicated that the on-demand operational model of UAM vehicles challenges traditional maintenance paradigms due to the lack of standardized components and the variability in operational conditions (Sieb, 2023; IFAR, 2023).

Unlike traditional air traffic that is confined to a few centralized airports, UAM operates through numerous dispersed vertiports, which inherently increases maintenance complexity and cost due to the need for frequent inspections and rapid response to potential failures. Moreover, high flight frequency and diverse operational environments necessitate non-destructive inspection methods that are both reliable and accessible to non-expert operators. Advancements in non-destructive evaluation techniques, such as ultrasonic and thermographic methods, have proven effective in early fault detection under variable operational conditions (Balasubramaniam, 2023; Liu, 2022).

Given these challenges, integrating artificial intelligence into maintenance systems offers a promising solution for early fault detection. However, direct data collection from UAM aircraft remains impractical due to their developmental stage and the high cost of inducing controlled defects. To overcome this, our study employs drone propellers—whose design and operational characteristics closely resemble those of UAM—to build a surrogate dataset for anomaly detection. Using this dataset, we developed an artificial intelligence model that monitors acoustic signals to identify subtle changes indicative of propeller defects, such as cracks. Recent advancements in acoustic signal processing and machine learning have further validated this approach by demonstrating high accuracy in detecting structural anomalies in aerospace components (Palanisamy, 2024; Sikdar, 2022). This approach not only demonstrates the feasibility of sound-based non-destructive testing but also lays the groundwork for its future application in enhancing UAM safety and maintenance efficiency.

## II. Related Works

### A. UAM (Urban Air Mobility)

UAM (Urban Air Mobility) is a new air traffic system that operates in conjunction with other modes of transportation, using environmentally friendly electric-powered vertical takeoff and landing aircraft (eVTOL) for purposes like passenger or cargo transport within urban areas. UAM can be used both within and outside urban areas for passenger and cargo transport, public purposes (such as emergency medical services), and tourism (Korea, 2021). As urbanization and population growth intensify in modern cities, UAM emerges as a new solution to address ground transportation challenges, drawing global attention. In particular, the United States and Europe are taking a keen interest, in establishing various regulations and procedures to lead the market.

The U.S. Federal Aviation Administration (FAA) defines UAM as "UAM is a safe and efficient aviation transportation system that uses highly automated aircraft to transport passengers or cargo at low altitudes in urban and suburban areas (Whitley, 2022)." Meanwhile, the European Union Aviation Safety Agency (EASA) describes UAM as, "UAM is a new aviation transportation system for passengers and cargo in densely built environments, made possible through eVTOL equipped with advanced technologies such as enhanced battery technology and electric propulsion (EASA, 2021)."

UAM aircraft are currently under development, with the goal of commercialization by 2025 after undergoing aircraft development and aviation certification procedures. So for now, most UAM aircraft manufacturers have released their aircraft designs, and the general trend and market direction can be anticipated.

The fuel for the UAM aircraft under development is electric, and they take the form of eVTOL, which lands and takes off vertically. They generate thrust through propellers, and as specified in Shim's report (Shim, 2021), depending on the propulsion method, they mainly showcase three aircraft designs: Multi Rotor, Lift & Cruise, and Tilt type.

1) **Multi-Rotor**: This design has multiple rotors and closely resembles drones in form.
2) **Lift & Cruise**: This design features both rotors and wings. During takeoff and landing, vertical rotors operate in a rotary wing fashion. However, during flight, horizontal rotors function as fixed wings.
3) **Tilt**: This term collectively refers to tilt rotors (rotors), tilt ducts (ducts), and tilt wings (wings). The distinction is based on which part rotates.

Meanwhile, UAM, given their nature resembling public transportation, typically have a shorter range of operation than conventional aircraft and therefore fly more frequently on average. As Yeon (Yeon et al., 2021) mentioned, when compared to traditional aircraft, UAM generates lower profitability per flight. This implies a need to reduce maintenance time and costs, aiming for quicker turnaround times.

In terms of maintenance locations, while traditional fixed-wing airlines possess hangars at each airport and their affiliated technicians directly maintain the aircraft, UAM operates a bit differently. UAM relies on "Vertiports," which serve as terminals or stops within urban areas, playing a similar role to traditional airports. Because these Vertiports are scattered throughout the city, it is challenging to centralize maintenance locations, parts, and staff. As a result, there's no need for separate hangars for UAM maintenance. Furthermore, it's essential to develop a system that allows even non-experts to identify aircraft defects quickly and easily.

### B. Anomaly Detection

Anomaly Detection is a field of artificial intelligence research that identifies abnormal data within a given dataset. It focuses on discovering data (abnormal data) that hasn't appeared yet but could potentially arise. In this context, "abnormal" refers to something that deviates from a normal state, differs from common experience or knowledge, is peculiar or unique, or is dubious and unidentifiable. This includes situations where the shape or state of images, sounds, or data deviates from what is typically considered normal.

Anomaly Detection serves as a mechanism to detect such abnormal data, mainly to identify machine malfunctions, recognize abnormalities in system logs, or verify abnormal states in images or videos. Essentially, it is used to detect conditions that are notably different while a standard state is maintained. This paper aims to study a method to detect abnormal conditions of UAM Propellers using anomaly detection artificial intelligence models.

According to Chalapathy et al (Chalapathy & Chawla, 2019), Anomaly Detection can be divided into Supervised Learning, Semi-supervised Learning, and Unsupervised Learning. Supervised learning is a method where one trains a classification model (Omar et al., 2013) using data labeled as normal and abnormal. When new data is input, the model estimates whether it's anomalous. Supervised learning has the advantage of guaranteed performance. However, due to the nature of fields where Anomaly Detection is used, it is difficult to acquire data on abnormal samples. Additionally, there is a significant data imbalance between normal and abnormal data. Fundamental classification models such as SVM, Decision Tree, and Deep Neural Network fall under this category.

On the other hand, the method most used for Anomaly Detection models is semi-supervised learning. In this method, the model is trained on one class of data (usually normal data) and determines the range for this trained data. If new incoming data falls outside this range, it is considered abnormal data. Semi-supervised learning has the advantage of showing substantial performance levels even when trained with one class and requires less data for training. However, there's a disadvantage: by determining the range for normal data, it might not reflect all the features of a small amount of abnormal data, leading to overfitting. Representative models for this approach include Deep SVDD (Zhou et al., 2021), Isolation Forest (Liu et al., 2008), and AutoEncoders (Baldi, 2012) that verify normality through data compression and restoration.

The unsupervised learning method detects anomalies based solely on the intrinsic properties of the data. Typically, data used for unsupervised learning isn't labeled. Instead, it's categorized based on its inherent attributes, and the model learns from these distinguished properties. However, in areas where Anomaly Detection is applied, most of the generated data is considered normal data. Thus, even without specific labeling, it can be assumed as normal. This approach resembles semi-supervised learning, which also trains on one class of data. A representative model for this is PCA (Hoffmann, 2007) (Principal Component Analysis), which identifies abnormal data during the process of dimension reduction and restoration.

**C. Acoustic Anomaly Detection**

Acoustic Anomaly Detection refers to a research method within Anomaly Detection that utilizes acoustic data. In other words, it's a research field that analyzes audio data signals recorded through microphones to identify abnormal data. There are various data analysis methods, such as Fourier transform and spectrum analysis. As Melchiorre's research (Melchiorre et al., 2023), anomaly detection is possible through the digitization of hearing, one of the human senses, which also implies that anomalies detected by humans can be verified and interpreted.

Using acoustic data for anomaly detection has advantages compared to using images or specialized chemical sensors. Namely, it leverages relatively low-cost acoustic sensors and has the advantage of collecting data without physical contact or interference with the target object. However, there are downsides. It's susceptible to ambient noise and interference, which can lead to data contamination. Additionally, acoustic sensors typically do not collect other situational information and only deal with sound signals, so the absence of visual or situational clue data may limit the complete understanding of the characteristics or causes of anomalies.

Nevertheless, with the continuous advancements in signal processing, machine learning, and sensor technology, many of these problems are continuously being resolved. Due to its convenient data collection method, it's one of the highlighted research areas in the Anomaly Detection field. Below are areas where Acoustic Anomaly Detection is being used.

1) **Industrial Monitoring**

   Acoustic anomaly detection can be used in industrial settings to monitor the condition of machinery and equipment (Kamat & Sugandhi, 2020; Purohit et al., 2019). By analyzing the acoustic signals under normal operational conditions, deviations or anomalies in sound patterns can indicate potential defects, malfunctions, or maintenance requirements.

2) **Surveillance and Security**

   Acoustic anomaly detection plays a crucial role in surveillance and security systems (Lane & Brodley, 1997; Vikram, 2020). By placing microphones or acoustic sensors in areas of interest, such as public spaces, transportation hubs, or critical infrastructures, suspicious or alarming sounds like gunshots, breaking glass, or explosions can be identified.

3) **Medical and Healthcare Applications**

   It aids in diagnosing and monitoring respiratory conditions (Han et al., 2021; Shvetsova et al., 2021). It's used to detect abnormal respiratory sounds like wheezing or crackling. Moreover, it helps identify abnormal heart sounds or analyze fetal heart sounds for prenatal monitoring.

## III. Method

**A. Methodology Overview and Comparative Analysis**

In this paper, due to the impossibility of collecting real data from UAM vehicles, we verified the feasibility of anomaly detection for UAM propellers using sound data from drone propellers, which are most similar in form to UAM (Geng et al., 2021). The similarities between the two aircraft are as follows:

1) Both drones and UAM obtain thrust and lift using propellers. They accelerate the air in a specific direction using the propeller to get thrust and obtain lift from the pressure difference above and below created by the propeller's rotation. The thrust and lift are controlled by adjusting the RPM of the propeller.

2) Both aircraft operate with electric motors. Since they can deliver maximum torque as soon as power is supplied, their response time is fast. Typically, they are quieter and have less vibration compared to internal combustion engines or other mechanical power sources. Furthermore, as both aircraft use a battery system for operating the electric motor, the voltage is not always consistent, hence the sound of the propeller operation also varies within a certain range.

A similar research approach can be found in the study by Koizumi et al. (Koizumi et al., 2019). In this research, they constructed the ToyADMOS Dataset, which collected sounds produced by various toy machines (trains, cars, helicopters, conveyor belts). Using this dataset, they developed an algorithm to detect various types of abnormal operations. While the sound of the toys collected might differ from real machinery in the details of the spectral shape, their time-frequency structure could be similar. By adjusting preprocessing parameters, it can be applied in the real world. Similarly, since drones and UAM share similarities in shape and propulsion methods, the frequency shape of the propeller sound is likely to be similar. With appropriate data preprocessing and model parameter adjustments, it can be inferred that they could be applicable in the real world.

Moreover, according to Wang et al. (Wang & Deng, 2018), in a Homogeneous domain setting, the source domain and the target domain exist in the same dimension, allowing for a one-step domain adaptation. Given the similarities in the structure of the two domains, it can be inferred that they are located within a Homogeneous domain. If we set the drone as the source domain and UAM as the target domain, domain adaptation might be feasible in the future.

Secondly, the core of this research is to detect defects in UAM propellers through non-destructive methods. In this paper, we focused on detecting defects through sound, among various non-destructive methods. When the propeller and motor operate, the propeller inevitably produces noise due to the vibration created when breaking through and accelerating the air. As proven in the experiment by Ghalamchi et al. (Ghalamchi & Mueller, 2018), if the shape of the propeller changes due to a defect, the pattern of the resulting vibration changes, subsequently altering the nature of the sound produced. Therefore, we adopted a non-destructive defect detection method through sound to research the anomaly detection methodology for UAM propellers. To further highlight the novel aspects of our approach, Table 1 below provides a direct comparison between existing methods and our proposed method.

**Table 1. Comparison of Existing Methods versus Proposed Method**

| Feature | Existing Methods | Proposed Method |
| --- | --- | --- |
| **Data Source** | Industrial machine sounds, toy models | Drone propeller sounds (ADCP dataset) |
| **Preprocessing Techniques** | FFT or STFT only | Combination of FFT & STFT for enhanced frequency-time representation |
| **Application Domain** | Industrial fault detection, manufacturing monitoring | UAM maintenance |
| **Microphone Angle Variability** | Limited angles, mostly fixed | Extensive evaluation across multiple angles (45°, 90°, 135°, etc.) |
| **Real-Time Feasibility** | Often constrained to controlled lab environments | Designed for potential real-world adaptation with varying noise conditions |
| **Anomaly Types Considered** | General industrial machine faults | Specific to UAM-related defects: propeller cracks (ripped, broken) |

### B. Research Procedure

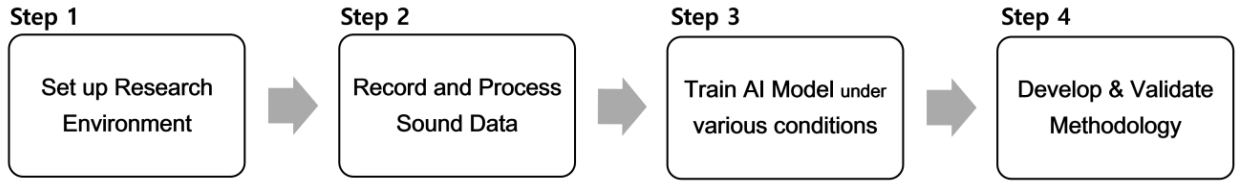

**Fig. 1 Research Procedure**

Based on the assumptions mentioned above, the research for detecting UAM propeller defects proceeded as shown in Fig. 1. First, to acquire the sound data of the drone's propeller, a research environment was set up, and data was generated through direct recording. Subsequently, the sound data was processed and preprocessed to train it with an artificial intelligence (AI) model. An appropriate model was selected for anomaly detection, and trained, and its performance was validated. By generating both normal and abnormal data on the propeller and validating the model for propeller defect detection, we could present a methodology to create an AI model capable of detecting defects in UAM propellers.

## IV. Environment Setup and Data Collection

To collect the sound data from the drone propeller, the propeller and motor were connected to a battery. To prevent vibrations caused by the thrust of the propeller, such as Soundararajan (Soundararajan & Jothi, 2021), it was securely mounted on a thrust bed and its operational sound was recorded with a microphone. According to Kurtz et al (Kurtz & Marte, 1970), as shown in Fig.2, due to the Doppler effect caused by the rotating propeller, the sound pressure level changes, and the shape of the vibration changes depending on the direction from the center of the vibration. Also, when the throttle changes, the RPM of the propeller changes, leading to a change in the intensity of the vibration. Given this, the direction of the propeller and power were adjusted to collect various forms of data.

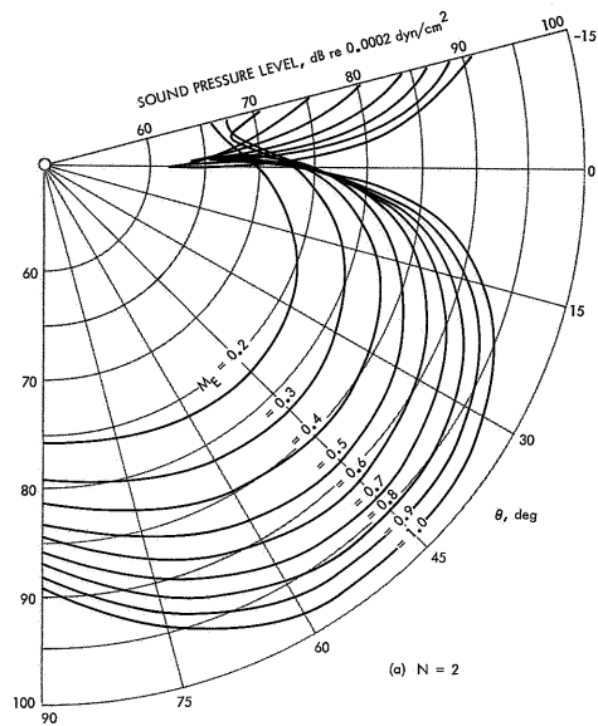

**Fig. 2 Variation of sound pressure according to the direction of propeller rotation** (Kurtz & Marte, 1970)

### A. Recording Environment and Setup

Facilities were available to securely mount the propeller and to ensure that no external noise would interfere with the recording, the wind tunnel laboratory of Konkuk University's heavy equipment research center was selected as the recording location. The wind tunnel laboratory operates using a closed-loop wind tunnel structure, where the flow is generated from a fan and supplied to the experimental model. The testing section where the model was placed is divided into the first section, where the flow circulated, and the second section, where the experimental subject was set up. For this study, the experiments were conducted in the second section, which has a space of 2.2m in width and 2.2m in height. While typical wind tunnel tests involved creating flow from a fan, for this study, to reduce interference from wind-related noise, no additional flow was generated. The interior of the wind tunnel lab was designed such that the first and second sections were connected. This design allowed the flow generated in the first section to circulate through the second section, where the test subject is placed. Hence, during the test, the wind generated by the propeller can exit through the first section, minimizing any reflective noise that might occur in a closed space and providing an optimal testing environment without interference from external noises.

(a) Placement of microphone   (b) Closed wind tunnel

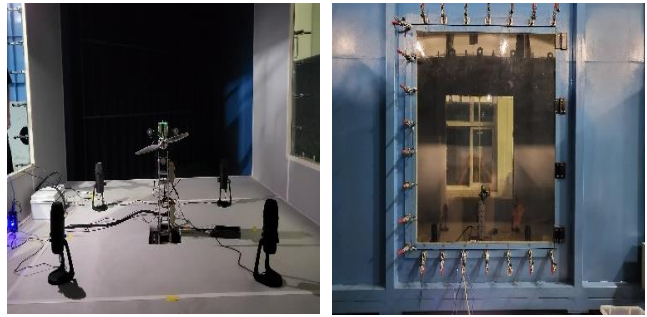

**Fig. 3  Recoding environment**

The experimental environment shown in Fig.3 was designed to avoid noise caused by reverberation in an enclosed space. To achieve this, sound-absorbing materials were attached to the walls. Inside the laboratory, the propeller and motor were mounted on a Thrust bed to prevent noise caused by vibrations, ensuring a stable experiment. Only a microphone and power equipment were installed where the propeller was operating. The control PC and the location of the experimenter were isolated by a glass door.

The microphone used was the BOYA USB condenser microphone, model BY-PM700. This microphone has a resolution of 16bit / 48kHz and can record in omnidirectional, unidirectional, stereo, and bipolar patterns. The recordable frequency range is 20 – 20,000Hz, and its sensitivity is -45±3dB@1kHz. During experiments, the microphone was set to omnidirectional recording to prevent the proximity effect (Milanova & Milanov, 2001) that could occur due to the distance between the microphone and the sound source. To evaluate the effect of microphone placement on anomaly detection accuracy, recordings were conducted at multiple angles (45°, 90°, 135°, 225°, 270°, and 315°) relative to the propeller. The propeller and microphone were kept 100cm apart. Also, to minimize the loss of acoustic data during recording, the recordings were made in the WAV format, which is a lossless and high-quality uncompressed file format. The chosen microphone distance and angle settings were selected based on prior studies on acoustic anomaly detection, ensuring a balance between capturing distinct propeller sound characteristics while minimizing environmental noise interference, as shown in Table 2.

**Table 2 Setting parameters for various data collection**

| Category | Normal | Abnormal |
|---|---|---|
| Numbers of Blades | 2 | 2 |
| Distance from the Propeller | 100cm | 100cm |
| Throttle | 20%, 20%, 40%, 50% | 20%, 20%, 40%, 50% |
| Direction of the Microphone | 45°, 90°, 135°, 225°, 270°, 315° | 45°, 90°, 135°, 225°, 270°, 315° |
| Crack | - | Ripped, Broken |

The propeller used was from the APC manufacturer, which is commonly mounted on unmanned aircraft. It's made of plastic material, with a length of 14 inches and a pitch of 10 inches. It's an electric propeller. The motor used was the KDE 5215XF-435 model from the KDE Direct company. While the RPM varies depending on the mounted propeller and battery, this model can output up to 11,350 rpm. Finally, the battery used was the PT-B10000-NSR35, which has an output of 14.8V and 35C.

**B. Dataset Composition**

Data was recorded by varying the Throttle and direction based on the variables in Table 2. In this context, Throttle is expressed as a percentage relative to the maximum performance of the available motor and battery. As the Throttle increases, the RPM also rises.

The dataset is broadly categorized into normal data and abnormal data.

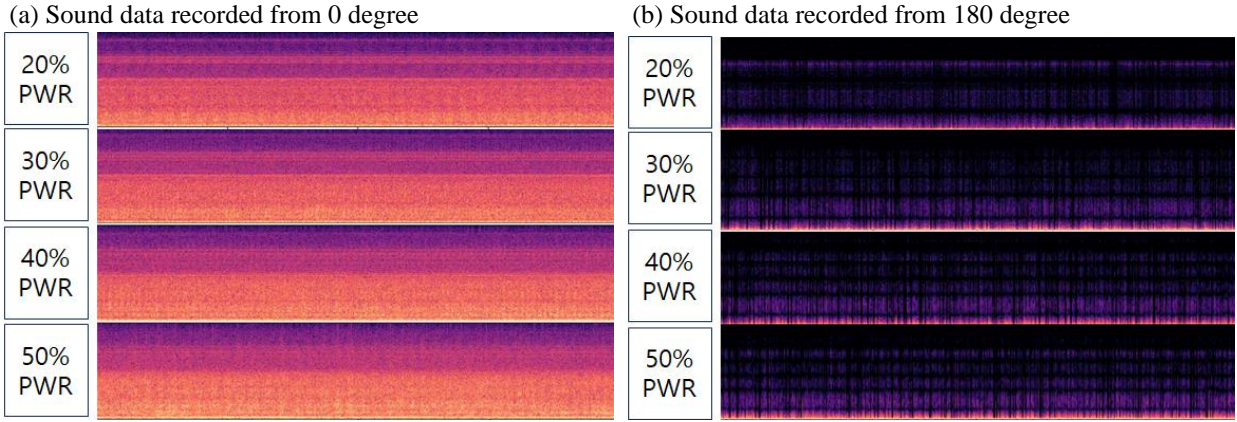

**Fig. 4  Sound data recorded from (a) 0 and (b) 180 degrees**

1) **Normal Data**

The sound of the propeller operating in its pristine condition was recorded by varying the direction of the microphone and the Throttle. Each variable was recorded for 30 minutes, and the recorded files were divided into 10-second segments. This produced 180 data samples for each variable. Therefore, for 24 variables, a total of 4,320 samples were obtained. However, among the data collection variables, when the direction of the microphone was at 0 degrees, as seen in Fig.4's first image, there wasn't a distinct variation observed. Also, at 180 degrees, due to the wind generated by the propeller blowing in the backward direction, the propeller sound was contaminated by the noise of the wind, as shown in Fig.4's second image, thus it was excluded.

2) **Abnormal Data**

The recording was conducted for all the variables of the normal data, with an addition of two types of Crack variables, resulting in a total of 48 variables. However, since the abnormal data was utilized as test data to determine the possibility of anomaly detection, it was not necessary to collect the same number of samples as the normal data. Therefore, the recording duration was reduced to 5 minutes for each variable, securing a total of 1,440 samples.

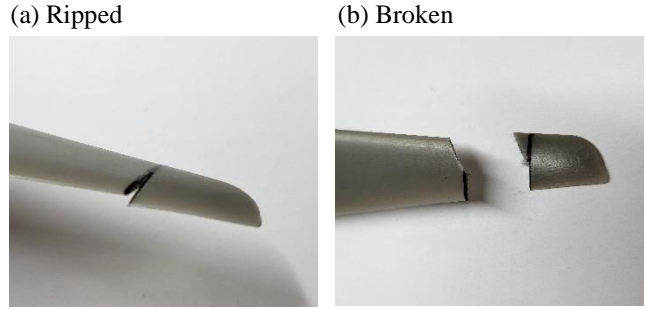

**Fig. 5 Types of Propeller Crack**

The Crack category is distinguished between 'Ripped' and 'Broken'. As shown in Fig. 5(a), a 'Ripped' is not a complete longitudinal damage of the propeller Blade. While its original shape is retained, visible gaps or crevices form along certain paths or lines. In this study, a condition where 10% of the entire propeller length (1.4 inches) has 50% damage is defined as the 'Ripped' type. 'Broken', as illustrated in the second image of Fig. 5(b), refers to the damage type where one side of the propeller Blade is completely broken, resulting in the Blade being divided into two separate pieces. This represents a damage type where 10% of the entire propeller length (1.4 inches) is fully damaged.

## V. Experiment

### A. Data Preprocessing

To analyze data recorded through a microphone, when loaded using sound data analysis tools in Python libraries (such as Python, librosa, and scipy), it is represented as a digital audio signal, an array of numeric values. The data type of these audio samples can be either 16-bit integers or 32-bit floating-point numbers. Each element in the array corresponds to an audio signal sample at a specific point in time. The Fourier Transform is the mathematical transformation technique that converts such audio signal samples from the time domain to the frequency domain. Through the Fourier Transform, a given signal can be decomposed and represented in terms of its various frequency components. It is used to extract frequency composition and amplitude information from the signal. This transformed data includes information about the magnitude and phase of frequency components, allowing for inspection of the original signal's frequency components, analysis for specific frequency bands, filtering, compression, and signal synthesis. The Fourier Transform is categorized into Fast Fourier Transform (FFT) and Short-Time Fourier Transform (STFT), and there are differences in the purpose and usage of these two transformations.

First, the Fast Fourier Transform (FFT) is a method that transforms a given signal across its entire time range all at once to obtain its complete frequency components (Cochran et al., 1967). Since this results in the loss of time information and only

(a) Normal data

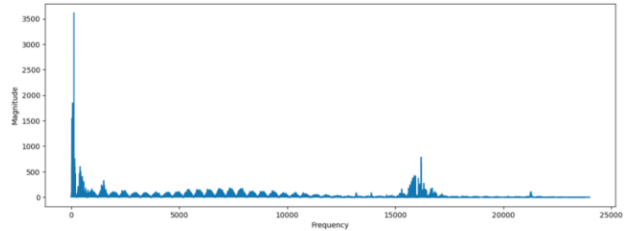

(b) Abnormal data

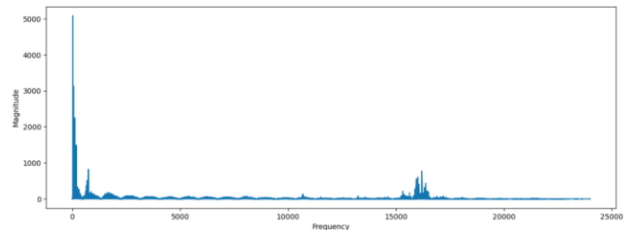

**Fig. 6  Data shape through Fast Fourier Transformation (FFT)**

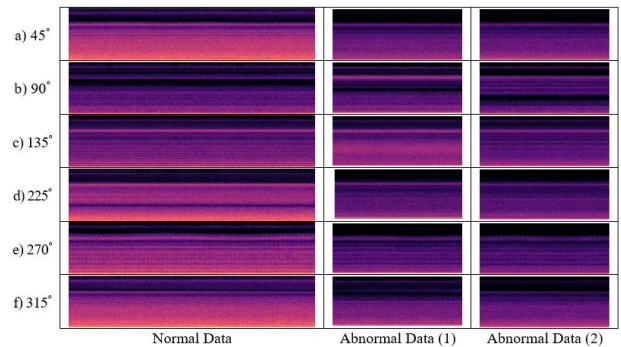

**Fig. 7  Data shape through Short Time Fourier Transformation (STFT)**

provides frequency information, it is suitable for analyzing the frequency characteristics of signals. On the other hand, the Short-Time Fourier Transform (STFT) is a transformation technique (Allen & Rabiner, 1977) designed to analyze a signal in both the time and frequency domains simultaneously. By dividing the time domain signal into fixed-sized windows and performing the Fourier Transform for each window, it preserves the temporal characteristics of the signal. STFT is mainly used in scenarios like analyzing frequency changes in musical signals over time or examining spectral changes in audio signals, especially when detailed information about time and frequency is required.

Fig. 6 illustrates the shape after applying the Fast Fourier Transform to the generated propeller sound data. The horizontal axis represents frequency information, and the vertical axis represents the magnitude of that frequency.

Observing the graph, we can see a distinct difference between normal and abnormal data in the frequency band from approximately 1,000Hz to 10,000Hz. In the normal data, we can observe a pattern where the values increase and decrease at regular frequency intervals. In contrast, this characteristic is significantly reduced in the abnormal data.

**Table 3 Setting parameters for various data collection**

| Category | | Train | Test | Category | | Train | Test |
|---|---|---|---|---|---|---|---|
| **Each Degree** | Normal | 640 | 120 | **Each Power** | Normal | 900 | 180 |
| | Abnormal 1 | - | 120 | | Abnormal 1 | - | 180 |
| | Abnormal 2 | - | 120 | | Abnormal 2 | - | 180 |
| **All Data** | Normal | 3600 | 720 | | | | |
| | Abnormal 1 | - | 720 | | | | |
| | Abnormal 2 | - | 720 | | | | |

On the other hand, Fig. 7 illustrates the visualization of the sound data after undergoing the Short-Time Fourier Transform. The horizontal axis represents time, while the vertical axis denotes frequency information. Moreover, the color of the graph signifies decibels, where the darker the color, the higher the decibel level. The graph in Fig.7 represents the normal and abnormal data (Abnormal Data (1): Ripped, Abnormal Data (2): Broken) according to the direction of the microphone. It is evident that the frequency characteristics and decibels differ depending on the direction. Additionally, since there are differences in frequency and decibel characteristics between normal and abnormal data in each direction, it is evident that the data is suitable for training artificial intelligence models for anomaly detection.

In this paper, we chose and applied either the Fast Fourier Transform (FFT) or the Short-Time Fourier Transform (STFT) based on which method was more suitable for the model using the collected data. The applied parameter values are as follows:

1) **Sample rate**: 48000
2) **n_fft**: 256 (length of the windowed signal after padding with zeros)
3) **n_mels**: 256 (number of Mel bands to generate)
4) **Window size**: 512
5) **Hope_length**: 256

The number of data used for model training and verification is shown in Table 3. During model construction, the recorded angles and Throttle Power were set as parameters, the model was trained, and the results were checked for each data type. Therefore, we trained and compared the results from datasets based on 6 recording angles, 4 Throttle Power levels, and 1 combined dataset, analyzing the training outcomes for a total of 11 datasets.

B. Model

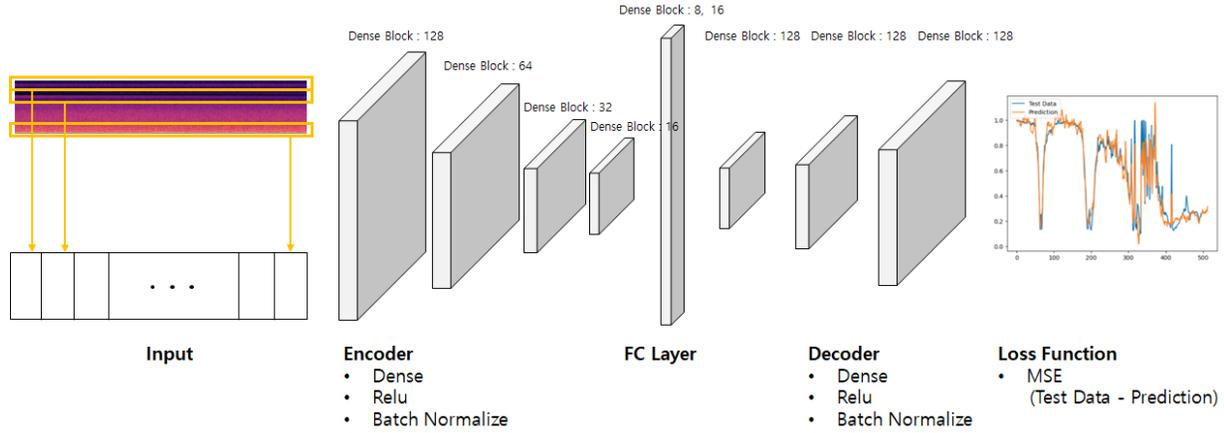

Fig. 8  The architecture of the AutoEncoder Model

Most of the fields where Anomaly Detection is used have limited access to anomalous data. Obtaining defective data from the UAM Propeller we are studying is also expected to be impossible in real-world scenarios. Reflecting this reality, this paper used a semi-supervised model that learns only from normal data to detect anomalies. Additionally, to verify the training status of the generated dataset, a basic anomaly detection model was utilized to establish a Baseline. The environment for model training is as follows:

1) **GPU**: NVIDIA GeForce RTX 4090

2) **CPU**: Intel Core i9-13900K 3GHz

3) **OS**: Ubuntu 20.04 LT

4) **Python**: 3.8

5) **Tensorflow**: 2.13

6) **Pytorch**: 2.0

The Baseline model in this paper employed AutoEncoder. AutoEncoder (Baldi, 2012) is an artificial neural network designed to learn an efficient or compressed representation of input data without explicit labels. It consists of an Encoder and Decoder. The Encoder reduces input data from a high dimension to a low dimension, projects data into a low-dimensional space, and the Decoder attempts to reconstruct the low-dimensional data back to a high-dimensional. The objective of AutoEncoder is to minimize the difference between input data and the reconstructed output data. Typically, it defines a loss function like mean squared error and performs learning by updating the weights and biases of the neural network using optimization algorithms like gradient descent.

The Encoder and Decoder are typically constructed by stacking various types of layers like FC Layer, Convolution Layer, LSTM Layer, etc. However, it is common that the number of neurons in the Encoding Layer is made fewer than the input data neurons. This ensures that during model training, the model can capture essential features or patterns of the data and discard unnecessary elements.

In this paper, we built an AutoEncoder model made up of FC (Fully Connected) Layer neurons. The data was preprocessed so that only data with decibels above a certain level could be emphasized, and then, as shown in the form of Input Data in Fig. 8, the data for each frequency was concatenated to change it into one dimension. Subsequently, the dimension was reduced through an FC Encoder that diminishes Dense Blocks to 128, 64, 32, and 16. After passing a Bottleneck with a Dense Block of 8, the data was reconstructed through a Decoder of 128. Adam was used as the optimizer, and the Mean Square Error was used as the loss function.

## C. Evaluation

The trained anomaly detection model is based on the reconstructed input data. Therefore, to detect anomalies, a process to derive an anomaly score is required. The anomaly score can be calculated through various methods, including Reconstruction error (Yamashita et al., 2006), Stochastic nearest neighbors (Idé et al., 2007), and the Depth of reached leaf in Isolation Forest (Liu et al., 2008). In this paper, the Reconstruction error, which is most used and intuitively calculated in the DNN AutoEncoder model, was used as the anomaly score. The anomaly score ($A_\theta(x_t)$) was calculated as the mean square error between the input data ($x_n$) and the prediction data ($f(x_n)$).

$$A_\theta(x_t) = \frac{1}{n}\sum_{i=1}^{n}\left(x_n - f(x_n)\right)^2 \tag{1}$$

To evaluate the anomaly detection model, two methods using the anomaly score can be considered. The first is the F1 score based on the confusion matrix, which is commonly used in classification models. As it considers both precision and recall, it reflects the characteristics of the real anomaly detection model where predicting the correct answer is important, but also emphasizes the precision of prediction. However, to calculate the F1 score in the anomaly detection model, the threshold is determined where normal and abnormal data can be best separated according to the distribution of anomaly scores, and the labeling test data is proceeded according to the anomaly scores separated by the threshold. Based on the separated anomaly scores determined by this threshold, correct labeling of the test data must be carried out. Hence, the evaluation value may vary depending on the threshold setting. Nevertheless, because it provides a balanced single evaluation metric by combining precision and recall, it can be considered a meaningful evaluation metric for anomaly detection models.

Next is the evaluation method based on the ROC AUC score according to the anomaly score. ROC AUC is a general evaluation method for anomaly detection models that evaluates the performance of distinguishing between normal and abnormal data at various thresholds (Iwata & Yamada, 2016). As it can be calculated with only the anomaly score and ground truth, it offers a more objective evaluation than the F1 score. A higher ROC AUC score indicates better differentiation between normal and abnormal data (Huang & Ling, 2005), suggesting that the model can more effectively differentiate between the two classes.

In our research, we compared both the F1 score and the ROC AUC score to determine which dataset model performs better.

## VI. Experiment Result

### A. Model Trained on Datasets Classified by Degree

**Table 4 Model Performance by recording angle**

| Degrees | F1 Score (Threshold) | | Roc AUC Score | |
|---|---|---|---|---|
| | **Ripped** | **Broken** | **Ripped** | **Broken** |
| **45°** | 82.84 (0.007) | 81.23 (0.006) | 87.10 | 87.41 |
| **90°** | 98.36 (0.0014) | 97.56 (0.0013) | 97.76 | 97.22 |
| **135°** | 100.00 (0.040) | 93.02 (0.017) | 100 | 96.88 |
| **225°** | 98.77 (0.016) | 97.56 (0.013) | 99.72 | 99.35 |
| **270°** | 79.45 (0.008) | 90.91 (0.008) | 88.19 | 96.56 |
| **315°** | 76.70 (0.011) | 84.55 (0.011) | 90.01 | 94.47 |
| **Average** | 89.35 | 90.80 | 97.80 | 95.32 |

Initially, we evaluated the performance by training an AutoEncoder on datasets classified based on the recording angle, one of the data parameters. During this process, we assessed the Abnormal data separately for both torn and cut data.

Table 4 displays the performance of models trained on datasets classified by the recording angle. A preliminary observation indicates that sounds from propellers when torn are best distinguished from regular noises when recorded at an angle of 135 degrees. On the contrary, recordings taken at 270° and 315° show relatively inferior performance. This suggests that the noise produced by a cracked propeller at these angles barely differs from the sound of a regular propeller.

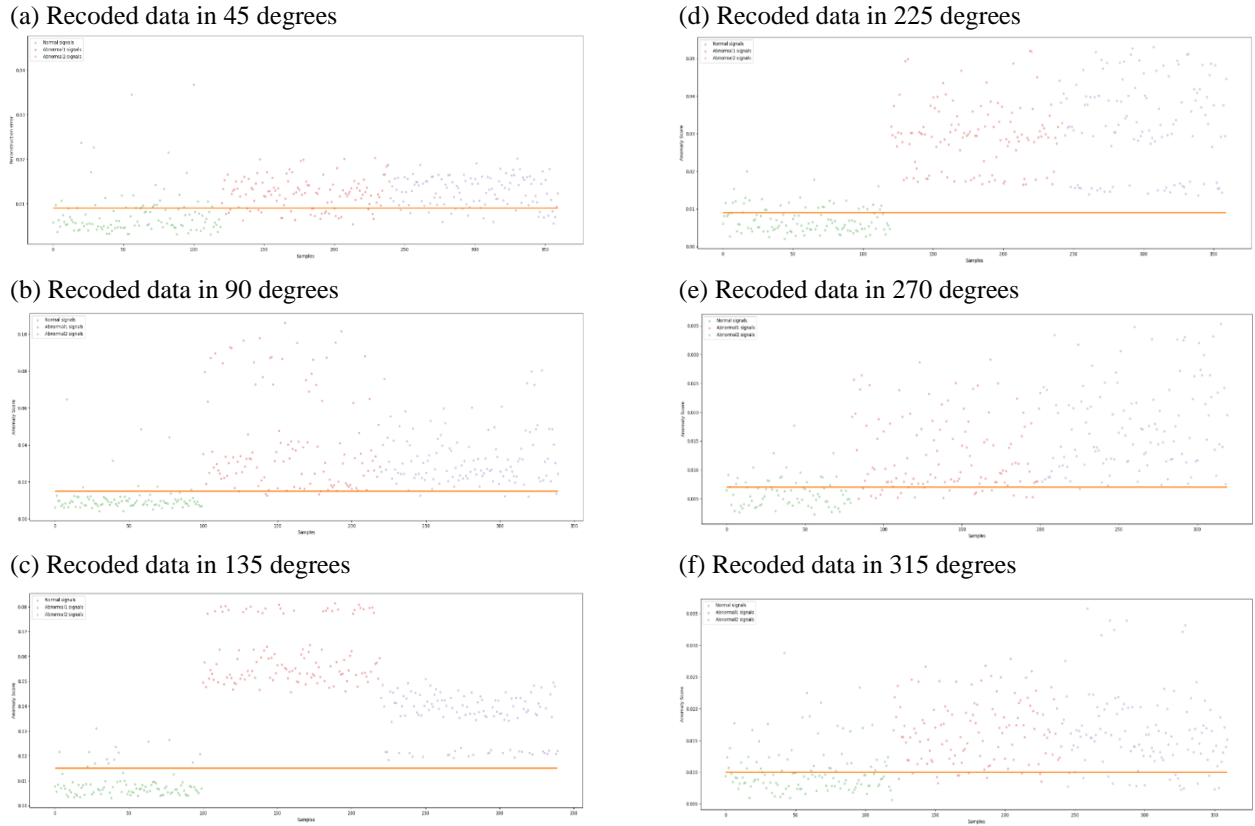

(a) Recoded data in 45 degrees

(b) Recoded data in 90 degrees

(c) Recoded data in 135 degrees

(d) Recoded data in 225 degrees

(e) Recoded data in 270 degrees

(f) Recoded data in 315 degrees

**Fig. 9 Distribution of Anomaly Score of datasets on each Throttle Power**

On the other hand, for Anormal data where part of the propeller is cut off, the sound is best detected when recorded at angles of 90° and 225°. Fig. 9 presents the distribution of anomaly scores as predicted by models trained on datasets corresponding to each angle. In each graph:

1) The leftmost 120 green dots represent Normal data.
2) The middle 120 red dots correspond to Abnormal data regarding ripping.
3) The rightmost 120 purple dots signify data associated with breaks.

This distribution helps to visualize how the model performs in differentiating between Normal and Abnormal sounds based on the recording angles, providing insights into the model's effectiveness and areas of potential improvement.

## B. Model Trained on Datasets Classified by Throttle Power.

**Table 5 Model Performance by All Data**

| Degrees | F1 Score (Threshold) | | Roc AUC Score | |
|---|---|---|---|---|
| | Ripped | Broken | Ripped | Broken |
| **20%** | 91.38 (0.010) | 90.43 (0.010) | 96.19 | 96.20 |
| **30%** | 90.81 (0.0011) | 89.32 (0.0011) | 96.93 | 96.52 |
| **40%** | 91.50 (0.013) | 92.82 (0.011) | 96.56 | 97.85 |
| **50%** | 71.84 (0.010) | 73.47 (0.009) | 71.43 | 76.98 |
| **Average** | 86.38 | 86.51 | 90.28 | 91.89 |

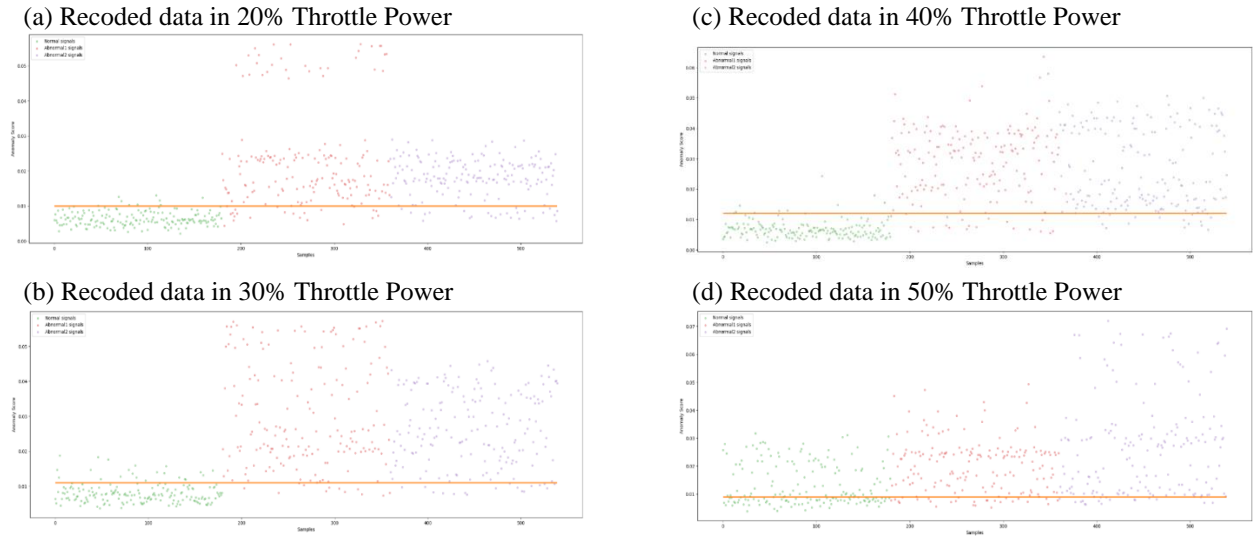

(a) Recoded data in 20% Throttle Power  (c) Recoded data in 40% Throttle Power

(b) Recoded data in 30% Throttle Power  (d) Recoded data in 50% Throttle Power

**Fig. 20. Distribution of Anomaly Score of datasets on each Throttle Power**

Next, we trained and evaluated the performance of individual datasets that were classified based on the data parameter, Throttle Power. In this case, as well, evaluations for Abnormal data were carried out separately for ripped data and broken data.

Table 5 displays the performance of models trained on datasets classified by Throttle Power. As the Throttle Power increases from 20% to 40%, we can observe a corresponding rise in model performance. However, there's a noticeable drop in performance as soon as it reaches 50%. This decline can be attributed to the increasing decibel level of the propeller noise and the enhanced presence of noise as the Power increases, making it challenging to capture the features of the data.

Fig. 20 represents the distribution of Anomaly scores predicted by models trained on datasets for each Throttle Power level. In each graph, the 180 green dots on the left represent Normal data, the 180 red dots in the middle denote Abnormal data regarding ripped, and the 180 purple dots on the right indicate data regarding break.

### C. Model Trained on All Data

**Table 6 Model Performance by All Data**

| F1 Score (Threshold) | | Roc AUC Score | |
|---|---|---|---|
| **Ripped** | **Broken** | **Ripped** | **Broken** |
| 79.52 (0.011) | 83.43 (0.012) | 84.83 | 89.22 |

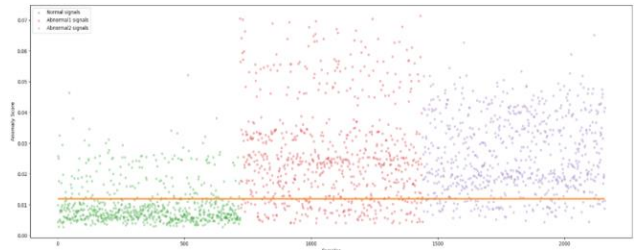

**Fig.21 Distribution of Anomaly Score of datasets on all data**

Lastly, we evaluated the performance of the model trained on all the generated data. As with previous evaluations, the Abnormal data was separately assessed for both ripped and broken data.

As looking at Table 6 and Fig. 21, the performance was found to be lower when the model was trained on all the data compared to when the datasets were trained separately. This is due to the angles (270°, 315°) and Throttle Power (50%) that, as previously observed, did not receive good evaluations and acted as outliers, dragging down the overall performance of the model.

However, the model overall demonstrated a performance of over 80% in both the F1 score and ROC AUC score. This confirms the ability to detect propeller cracks through sound using the dataset constructed in this paper. Furthermore, it was confirmed that there's potential for this method to be applied in the maintenance of Urban Air Mobility (UAM).

# VII. Discussion

To further interpret the experimental results and address key questions related to the methodology, this section discusses the technical contributions and limitations of our approach.

## A. Methodological Contributions

Our proposed method leverages a dual preprocessing strategy that integrates both FFT and STFT. This approach is significant because FFT provides a comprehensive frequency spectrum while STFT captures transient, time-dependent spectral features. By combining these techniques, our model yields richer spectral insights, which enhances anomaly detection accuracy compared to methods that use a single transformation. Additionally, the careful selection of evaluation metrics—using ROC AUC for its threshold-independent assessment and F1 score for optimal threshold tuning—further strengthens the reliability of our performance evaluation.

## B. Limitations and Challenges

Despite promising performance under controlled conditions, several challenges remain for real-world application. Ambient noise in operational environments may degrade performance, and the ADCP dataset's controlled nature does not fully capture such complexities. Moreover, our experiments revealed that microphone placement significantly impacts detection accuracy, indicating the need for additional testing in dynamic settings. Scalability is another concern, as current experiments focus on a single propeller; transitioning to multi-propeller systems will require advanced sensor fusion techniques. Finally, differences in acoustic features—such as those between electric motor-driven and internal combustion engine-driven propellers—necessitate tailored preprocessing and threshold settings.

## C. Future Directions and Broader Implications

Future work will address the outlined limitations by exploring noise robustness and domain adaptation techniques (e.g., adaptive filtering and deep learning-based noise reduction), as well as dynamic threshold adaptation methods. Enhancing scalability through multi-microphone arrays and advanced data fusion strategies is also a key priority. Beyond UAM maintenance, the proposed methodology shows potential for early anomaly detection in various domains, including automotive maintenance, wind energy systems, and industrial machinery, where timely fault detection is critical.

## VIII. Conclusion

In the continuously evolving UAM market, ensuring aircraft safety through thorough preventive maintenance is of paramount importance, while economic efficiency must not be overlooked. The integration of artificial intelligence into maintenance practices provides a promising solution to balance these needs. In this paper, we presented a novel methodology for detecting propeller defects using a drone propeller sound dataset, marking the initiation of AI-assisted maintenance for UAM systems.

By constructing a comprehensive dataset from drone propeller sounds and employing a dual preprocessing strategy (FFT and STFT), we demonstrated the potential for effective anomaly detection. Our experimental results, as indicated by robust F1 scores and ROC AUC metrics, confirm that the proposed approach reliably distinguishes between normal and defective propeller conditions under controlled conditions.

The controlled experimental setup allowed us to capture high-quality, noise-mitigated data by simulating various defect scenarios (10% ripped or broken) and varying key parameters such as microphone angle and throttle power. However, since the data were collected in a closed space, real-world UAM operations, which are subject to ambient urban noise, may present additional challenges. Future research will focus on recording propeller sounds in open environments and refining the dataset to include defects of less than 10%, ultimately aiming to detect even minor anomalies.

Overall, our study provides a solid foundation for the application of AI-based anomaly detection in UAM maintenance, with broader implications for automotive maintenance, wind energy systems, and industrial machinery. Addressing the limitations identified in controlled settings will be crucial for translating this approach to real-world applications and further enhancing its robustness and scalability.